\documentstyle{mn}

\input{epsf}

\def\osse{{\it OSSE~}}
\def\comptel{{\it COMPTEL~}}
\def\cgro{{\it CGRO~}}
\def\egret{{\it EGRET~}}
\def\rosat{{\it ROSAT~}}
\def\cosb{{\it COS-B~}}

\def\ee{$e^\pm$~}
\def\ergs{~{\rm erg}~{\rm s}^{-1} }
\def\ergscm2{~{\rm erg}~{\rm s}^{-1}~{\rm cm}^{-2} }
\def\MeV{~\rm{MeV}}
\def\Lg{$L_\gamma$}
\def\edot{$L_{\rm sd}$}

\newbox\grsign \setbox\grsign=\hbox{$>$} \newdimen\grdimen \grdimen=\ht\grsign
\newbox\simlessbox \newbox\simgreatbox \newbox\simpropbox
\setbox\simgreatbox=\hbox{\raise.5ex\hbox{$>$}\llap
     {\lower.5ex\hbox{$\sim$}}}\ht1=\grdimen\dp1=0pt
\setbox\simlessbox=\hbox{\raise.5ex\hbox{$<$}\llap
     {\lower.5ex\hbox{$\sim$}}}\ht2=\grdimen\dp2=0pt
\setbox\simpropbox=\hbox{\raise.5ex\hbox{$\propto$}\llap
     {\lower.5ex\hbox{$\sim$}}}\ht2=\grdimen\dp2=0pt
\def\simgreat{\mathrel{\copy\simgreatbox}}
\def\simless{\mathrel{\copy\simlessbox}}

\title[Gamma-ray pulsars]
{The efficiency of gamma-ray emission from pulsars}
\author[B. Rudak \& J. Dyks]
       {Bronis{\l}aw Rudak$^1$, Jaros{\l}aw Dyks$^2$ \\
$^1$N. Copernicus Astronomical Center, Rabia\'nska 8, 87-100 Toru\'n,
Poland\\ e-mail: bronek@camk.edu.pl\\ 
$^2$Dept. of Physics and Astronomy, UMK, Grudzi{\c a}dzka 5/7, 87-100 Toru\'n, Poland\\
e-mail: jinx@astri.uni.torun.pl }

\date{in the press}
\pagerange{\pageref{firstpage}--\pageref{lastpage}}

\begin{document}

\maketitle

\label{firstpage}

\begin{abstract}
We present a modified scenario of gamma-ray
emission from pulsars within the framework of 
polar cap models. 
Our model incorporates possible acceleration of
electron-positron pairs created in magnetospheres, and their subsequent
contribution to gamma-ray luminosity \Lg. It also reproduces the empirical trend
in \Lg~ for seven pulsars detected with {\it Compton Gamma Ray Observatory} (\cgro) experiments.
At the same time it avoids basic difficulties
(Nel et al. 1996, Arons 1996) faced by theoretical models
when confronted with observational constraints. 

We show that the classical and millisecond pulsars form 
two distinct branches in the \Lg~-- \edot~diagram (where \edot~is the spin-down luminosity).
In particular, we explain why the millisecond pulsar J0437-4715 has not been detected 
with any of the \cgro
instruments despite its very high position in the  ranking list
of spin-down fluxes (i.e. $L_{\rm sd}/D^2$, where
$D$ is a distance).
The gamma-ray luminosity predicted for this particular object is  
about one order of magnitude below the upper limit set by \egret.

\end{abstract}

\begin{keywords}
gamma-rays: theory, observations -- pulsars: general
\end{keywords} 

\section{INTRODUCTION}
\label{s:intro}
Numerous models of pulsars (see Michel 1991 for the review) proposed over the last
three decades tried to make specific predictions about emission of
gamma-rays and X-rays. Gamma-rays are particularly important
as a direct signature of
basic non-thermal processes in pulsar magnetospheres,
and potentially should help to discriminate among different models. Interpreting 
gamma-rays should also be less ambiguous compared to X-rays. In
the latter case, especially for objects younger than $10^6$ yr, contributions
from initial cooling, internal friction, etc.
of unknown magnitude 
may dominate the total X-ray emission.

There are seven positive detections of pulsars by \cgro (see Table 1), i.e. less than
one per cent of all pulsars known to date.
In all cases 
the sources
had been identified by virtue of gamma-ray flux modulations with 
previously known $P$.
Crab and Vela are the only pulsars seen by three of \cgro detectors.

\begin{table*}
\quad\quad\quad\label{t:par}
%\centering
\caption{Gamma-ray luminosities for pulsars detected with \cgro
(in log of $[\ergs ]$). Beaming angle of emission $\Omega_\gamma = 1 {\rm ~sr}$ was assumed.}
   
\begin{tabular}{lcccccccccc}
\hline
PSR      &        & log \edot & \egret & \comptel & \osse  & Refs. & `total'\\
\hline
B0531$+$21 & Crab   & $38.65$ & $34.6$ & $35.0$   & $35.1$ & 1,2,3 & $35.42$    \\
B1509$-$58 &        & $37.25$ & -- & -- & $34.4$ & -,-,3 & $34.40$   \\
B0833$-$45 & Vela   & $36.84$ & $34.2$ & $33.5$ & $31.3$ & 1,2,3 & $34.28$    \\
B1951$+$32 &        & $36.57$ & $34.1$ & $34.1$ & -- & 1,4,- & $34.40$          \\
B1706$-$44 &        & $36.53$ & $34.4$ & -- & -- & 1,-,- & $34.40$            \\
J0633$+$1746 & Geminga$^*$ & $34.51$ & $32.9$ & $+$ & -- & 1,5,- & $32.90$        \\
B1055$-$52 &        & $34.48$ & $33.4$ & -- & -- & 6,-,- & $33.40$         \\
\hline \\
\end{tabular}\\
$^*$at the distance of 157 pc. \\ 
Refs. to flux values:\\
1) Ramanamurthy et al.(1995), 
2) Carrami{\~n}ana et al.(1995),
3) Schroeder et al.(1995),\\
4) Kuiper et al.(1996a), 
5) Kuiper et al.(1996b),
6) Fierro (1995)\\

\end{table*}

The \egret data became so far the only ground for testing theoretical
models. The latest critical review of three models of gamma-ray emission
(polar cap models by Harding 1981, Dermer \& Sturner 1994, 
and outer gap model of Yadigaroglu \& Romani 1995)
comes from Nel et al.(1996). They conclude that
none of the models fits observations satisfactorily for two major reasons. 
First, 
the confrontation of polar cap models (Harding 1981, and Dermer \& Sturner 1994)
with the observations leads to a~relation
\begin{equation}
L_\gamma({\rm observ.}) \propto L_\gamma^\alpha({\rm model}),  
 \label{e1}
\end{equation}
with $\alpha \simeq 0.6$ and 0.5, respectively, instead of $\alpha = 1$. 
Second, there have always been some cases amongst $\sim 350$ \egret upper limits, apparently
contradicting predictions of \Lg~ made by the models. The troublesome limits come usually from the pulsars 
B1509-58,
B1046-58, B0656+14, B1929+10, B0950+08, as well as from the millisecond object J0437-4715.

\begin{figure}
%\centering
%\vspace{14 cm}
\begin{center}
\leavevmode
\epsfxsize=8.4cm \epsfbox{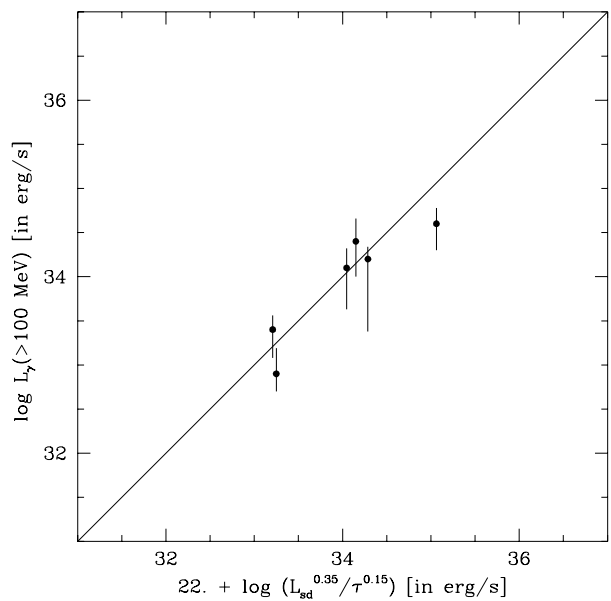}
\end{center}
\caption{ Gamma-ray luminosity above $100$ MeV for 6 \egret sources
is plotted against predictions of the polar cap model of Harding (1981).
The normalization factor is arbitrary; $\tau = P/2\dot P$ is a characteristic age expressed in [yr].
(Note: The Crab pulsar is the first dot from the right.) Bars indicate uncertainties in \Lg~
arising from uncertainties in distance and flux. 
In the case of Vela (next to Crab) the lower
distance limit was extended down to 200~pc 
to conform to recent estimates
based on X-ray and gamma-ray observations
of the Vela SNR ($400 \pm 200$pc after Aschenbach et al.1995 
and $\simless 350\,$pc after Oberlack et al.1994, respectively).
 }
\label{fig:fig1}
 \end{figure}

Of these two problems raised by Nel et al.(1996), the latter is more severe in our opinion. 
The former problem may be
solved to some degree by updating parameters used for two objects at the low-luminosity domain, i.e. B1055-52
and Geminga. After Thompson et al.(1994), Nel et al. used for B1055-52 the spectral index $\gamma = 1.18$ 
determined by Fierro et al. (1993) from the first three viewing periods of \egret.
However, a substantially higher value, $\gamma = 1.59$, based on the data from 10 viewing periods 
became recently available (Fierro 1995).
This value of the spectral slope reduces \Lg~of B1055-52 by a factor of $\sim 4.$
(We shall discuss other consequences of the steeper spectral
slope for B1055-52 in Section 3). Further reduction of \Lg~in the case of B1055-52 is possible by 
lowering distance
$D$ to the source. The argument for lowering the distance  (the usually assumed
value is $1.53\,$kpc, after the model of Taylor \& Cordes 1993) may
come from \rosat PSPC observations. If the bulk of the (presumably) thermal
X-ray emission from B1055-52 is due to
initial cooling then the inferred radius for the neutron star at $D = 1.53\,$kpc exceeds
$30\,$km ({\"O}gelman 1995) - a value hardly acceptable from the theoretical point of view.
In the case of Geminga, its inferred luminosity drops by a factor of $\sim 2.8$ relative 
to the value used by Nel et al. due to a newly determined HST parallax distance of $157\,$pc 
(Caraveo et al. 1996), instead of 
$D = 250\,$pc.
Fig. \ref{fig:fig1} presents the inferred \egret luminosities versus model predictions of Harding (1981),
with Geminga, and B1055-52 points corrected for $D = 157\,$pc, and $\gamma = 1.59$, respectively. 
Generally,
the agreement between theory and observations looks quite satisfactory, even though 
the Crab pulsar visibly deviates from the trend predicted by Harding (1981).

Arons (1996) presented a simple argument against all models which offer a functional relation
for \Lg~ roughly similar to
$L_\gamma \propto B/P^2$, by plotting the voltage ($\sim B/P^2$)
available for
particles, against the gamma-ray luminosity $L_\gamma(>100\MeV)$ for six \egret pulsars. An extrapolation of
the resulting trend towards
low values of $B/P^2$ leads to \Lg~ exceeding \edot~ below $\sim 10^{14}\,$Volts. Since \Lg~must never exceed \edot, 
this special value of voltage defines then `an empirical gamma-ray death line'
on $P-\dot P$ diagram (or on $Voltage-P$
diagram, as originally presented by Arons). It is hard to reconcile such a~line
with the observed death line for
radio emission, which corresponds to $\sim 10^{12}\,$Volts. The effect was known
already earlier, 
when polar cap models of Buccheri et al.(1978) and Harding (1981) had been introduced. 
In the latter model, for a pulsar with $\dot P \approx 10^{-15}$s~s$^{-1}$, its \Lg~ would reach \edot~
at a characteristic age equal to $3 \times 10^7$ years (Harding 1981).

%Both popular polar cap models 
%(Harding 1981, and Dermer \& Sturner 1994) suffer from this effect.

Below we propose a polar cap model
which is free of the two problems discussed above. This model reproduces gamma-ray luminosities 
inferred for seven
observed pulsars. At the same time
it avoids the problem of the empirical gamma-ray
death line of Arons (1996), 
and it relaxes the upper limits' constraints of 
Nel et al. (1996), especially for old classical pulsars, and
millisecond pulsars. In Section 2 we start with recalling the model for total power contained
in outflowing particles, which refers directly to the relation analysed by Arons (1996). Then we present arguments 
for summing all available \cgro data in order to get a better start for a modified polar cap model
rather than using the \egret data alone. Section 3 contains a description of our model, and its reference
to existing information on gamma-rays from pulsars. Summary and comments are in
Section 4; it contains also the ranking of pulsars with the highest gamma-ray fluxes resulting
from our model.

\section{SIMPLE MODELS VS. {\it CGRO} DATA}
\label{s:section2}

Though \egret has been far more successful than other
\cgro instruments in detecting pulsars,
this does not
mean that pulsars' gamma-ray emission above 100~MeV
dominates energetically over gamma-ray emission from lower energy bands. 
In the case of the Crab pulsar most of the energy output occurs within
the \comptel and \osse energy ranges (e.g. Fierro 1995). 
In the extreme case of B1509-58 there is
only \osse detection. Early positive reports from the \comptel team (Carrami{\~n}ana 
et al. 1995) have not been confirmed (Kuiper 1996), and \egret put only upper
limits for the source (Thompson et al. 1994).
Luminosities listed in
Table~1 were inferred from phase-averaged fluxes assuming beaming solid angle
of gamma-ray emission $\Omega_\gamma$ equal 1 steradian. For 3 objects (Crab, Vela, and B1951+32) detected with
more than one instrument, we can also construct a sum of inferred luminosities (ignoring
possible changes of a beaming angle with energy range), to get an estimate of `total' gamma-ray luminosity.
For the remaining 4 pulsars we will use their luminosities in the \egret energy range as `total'. 
The values of `total' gamma-ray luminosities are shown in the last colum of Table 1.

Let us now compare these `total' gamma-ray luminosities
with the simplest possible phenomenological polar cap model. 
According to this model,
gamma-ray luminosity \Lg~is proportional to a power contained in outflowing primary electrons 
$L_{\rm particles}$,
which in turn is proportional to a product of 
primary electron
energy $E_0$, a surface area
of canonical polar cap $A_{\rm pc} \approx \pi R_{\rm pc}^2 \propto 1/P$, and a Goldreich-Julian
flux $\dot n_{\rm GJ} \propto B/P$ (Goldreich \& Julian 1969) of outflowing primary electrons:
\begin{equation}
L_\gamma \propto L_{\rm particles} \propto E_0 \cdot \dot n_{\rm GJ} \cdot A_{\rm pc}. 
 \label{e2}
\end{equation}
Assuming $E_0 = {\rm const.}$ for all objects one obtains
\begin{equation}
L_\gamma \propto L_{\rm particles} \propto B/P^2 \propto L_{\rm sd}^{1/2}. 
 \label{e2a}
\end{equation}
(Note: The model for $L_{\rm particles}$  from equation (\ref{e2a}) was actually a starting point 
in the work of Harding 1981, who assumed $E_0 = 10^{13}$eV and whose objective
was to find a prescription for $L_\gamma(>100\MeV)$. 
We shall return to this point in the next section.)

\begin{figure}
%\centering
%\vspace{14 cm}
\begin{center}
\leavevmode
\epsfxsize=8.4cm \epsfbox{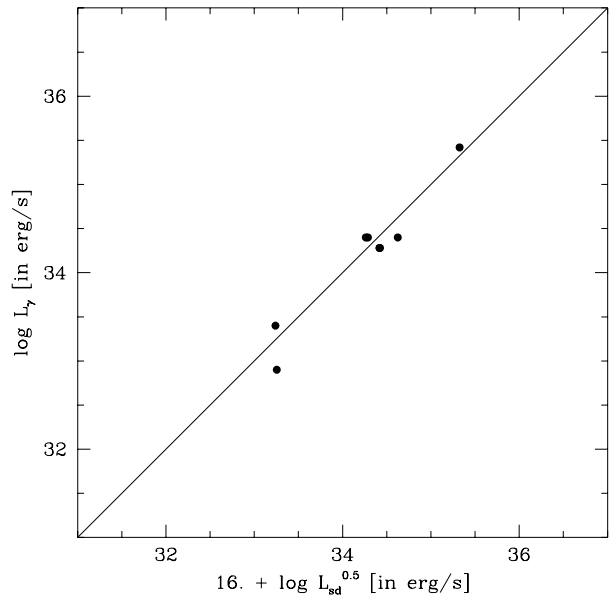}
\end{center}
\caption{`Total' gamma-ray luminosity $L_\gamma$ for all seven \cgro sources,
inferred from \egret, \comptel, and \osse
observations (wherever available - see Table 1.) are compared
with a simple model (equation \ref{e2a}) discussed in Section 2. 
(Note: B1951+32 and B1706-44 practically coincide.
The dot lying at the upper right part
of the diagonal represents the Crab pulsar.)
 }
 \label{fig:fig2}
 \end{figure}

Figure \ref{fig:fig2} shows how the relation of equation (\ref{e2a}) compares with observations. 
The normalization factor ($C = 10^{16}$) has been chosen
to obtain the best `by eye' fit. The overall agreement looks quite impressive. 
There is a substantial improvement for Crab comparing to Figure \ref{fig:fig1} due to 
significant contributions from
\osse and \comptel data.
Moreover, B1509-58 adds up smoothly to six \egret pulsars. The same functional relation,
but for \egret points only, has been 
considered by Arons (1996) (see previous section).

The relation $L_\gamma = 10^{16} L_{\rm sd}^{1/2}[\ergs]$ presented in
Fig.\ref{fig:fig2},  
cannot hold for all radio pulsars. It leads formally to 
\Lg~reaching
\edot~at $10^{32}\ergs$ (which corresponds to the Arons' empirical gamma-ray death line),  
whereas pulsars are observed down to $L_{\rm sd}\simeq 10^{30}\ergs$. 
Clearly, pulsar models which predict \Lg~as a simple 
combination of $B$ and $P$, require some revision.

\section{HOW DO ELECTRON--POSITRON PAIRS CONTRIBUTE TO GAMMA-RAYS}
\label{s:rq}
According to the model of Daugherty \& Harding 1982 (DH82)
primary electrons are 
accelerated
along open magnetic field lines
to high energies ($\sim 10^{13}$eV) due to 
rotation-induced electric field. The model assumes a dipolar structure of the magnetic field. 
Curvature photons emitted by primary 
electrons are absorbed by magnetic field with subsequent creation
of electron-positron pairs (Sturrock pairs). These pairs cool off instantly
via synchrotron radiation. Synchrotron photons may lead to further
pair creation. Electromagnetic cascades propagating
in pulsar's magnetic field may be very rich, with several 
subsequent generations
of pairs and photons.

Numerical treatment of
electromagnetic cascades initiated by primary electrons
above polar cap, and propagating across the magnetosphere 
(DH82) does not include effects of 
possible acceleration of Sturrock pairs. The only contribution from pairs to gamma-rays
taken into account is due to synchrotron emission of created pairs. The pairs themselves
do not accelerate, and subsequently - do not contribute to the curvature radiation.
Such simplification is usually justified by arguing that an appearence of conductive plasma
above some height effectively leads to a~screening of electric field
parallel to local magnetic field lines. If, however, the density of created pairs is lower
than the local corotation plasma density, the electrons from pairs will be subject to
further acceleration (whereas positrons will be decelerated; eventually some of them
will be stopped and reversed towards the stellar surface). 
In the context of polar cap models
developed by Daugherty \& Harding
(1982, 1994, 1996) it became clear that potential contribution to gamma-rays
from pairs might be necessary to account for the observed gamma-ray fluxes .  

If these secondary particles
are indeed subject to effective acceleration at significant altitudes above
polar cap surface, e.g. at heights of several NS radii (Daugherty \& Harding 1996),
the resulting beaming angles
of gamma-ray emission $\Omega_\gamma$ will be wider than those measured at
the polar cap surface. The requirement for ``nearly aligned rotators" (Dermer \& Sturner 1994, 
Daugherty \& Harding 1994),
necessary to explain large duty cycles of gamma-ray emission, might be then relaxed.
[Note: Relaxing the assumption about small inclination angles between the rotation
and magnetic axes may be inevitable
on observational grounds. Inclination estimates carried out by Lyne \& Manchester (1988),
and Rankin (1990) show that in many pulsars inclination angles are large indeed. 
However, mutual comparison of these results shows that they
are in agreement only for small inclination angles ($\simless 40\deg$), and there is no correlation
if either estimate is larger than this value (Miller \& Hamilton 1993).]
Hereafter we will assume that for all pulsars $\Omega_\gamma = 1 {\rm sr}$ 
(corresponding to opening angles of $\sim 30$ degrees).

Suppose, that the secondary particles, which we assume to
be created as described in the model of DH82, do participate in the gamma-ray production
similarly as the primary electrons do.
In the spirit of equations (\ref{e2}) and (\ref{e2a}), we propose a following prescription for \Lg:
\begin{equation}
L_\gamma = C \cdot n_{\pm} \cdot E_{\pm} \cdot L_{\rm sd}^{1/2}, 
 \label{e3}
\end{equation}
where $n_{\pm}$ is a number of created pairs \ee per primary electron,
and $E_{\pm}$ is a characteristic energy attained by particles due to acceleration. 
The normalization constant $C$ will be determined by fitting the observations.
We will assume
that $E_{\pm}$ achieved by secondary particles is similar to the energy attained by
primary electrons $E_0$, i.e. $E_{\pm} \simeq E_0$.

We begin with making a choice for the value of $E_0$, since it is
this parameter which, along with $B$ and $P$, will determine the number
of pairs $n_{\pm}$ created per primary electron. 
The analytical fit of Harding (1981) for \Lg~above 100 MeV, which 
gained so much popularity in testing her polar cap model against \egret observations was obtained
from numerical simulations performed for a fixed value of primary electron energy 
$E_0 = 10^{13}{\rm eV}$. Originally, this energy has been chosen to make the best spectral fits
above 100 MeV for \cosb data of Crab and Vela. However, such assumption about $E_0$ is not possible
throughout the entire pulsar's lifetime
because the energy of outgoing electrons
is subject to twofold limitation (e.g. Sturrock 1971): there 
is an absolute upper limit 
\begin{equation}
E_{\rm W} = 1.2 \times 10^7 \, B_{12} P^{-2} \,[\MeV]
\end{equation}
($B_{12} = B/10^{12}\,$G, period $P$ is in [s])
set
by potential drop across the polar cap, and a maximum value 
\begin{equation}
E_{\rm max} = 4.6\times 10^7 \, B_{12}^{1/4} P^{-1/8}\, [\MeV] 
\end{equation}
set up by
curvature cooling in purely dipolar magnetic field. 
The energy $E_0$ of primary electrons must conform to
one of these two limits (whichever comes first) as the pulsar is slowing down, since
both $E_{\rm W}$ and $E_{\rm max}$ decrease as $P$ increases.
In consequence, 
the prescription for $L_\gamma(>100\,\MeV)$ of Harding (1981) should not be
treated as accurate everywhere.
 
Moreover, accelerating electrons should cross threshold
energy 
\begin{equation}
E_{\rm min} = 1.2 \times 10^7 \, B_{12}^{-1/3} P^{1/3} [\MeV]
\end{equation}
required for pair creation. 
Whenever $E_{\rm W}$ falls below
$E_{\rm min}$, a classical pulsar crosses the well known deathline, and enters `a graveyard for pulsars'.
Another deathline occurs for millisecond pulsars even earlier, when $E_{\rm max}$ falls below
$E_{\rm min}$ (Rudak \& Ritter 1994). 

An ultrarelativistic primary electron, sliding along curved magnetic field line, emits curvature
photons which in turn may be converted into \ee pairs. The total number of pairs created per
electron depends
on the local component of magnetic field
$B$ perpendicular to the
direction of propagation of the photon, on the energy of the electron $E_0$, and on the curvature of magnetic
field lines. 
As the primary electron accelerates, its energy $E_0$ will cross the threshold value
$E_{\rm min}$,
thus triggering creation of a first pair.
The number of created pairs quickly increases as $E_0$ exceeds $E_{\rm min}$.
At some point
it becomes high enough to make further acceleration of electrons less effective due to screening effects. 
Instead of fixing $E_0$ at some specific value we assume thus that
electrons may be accelerated up to an energy $E_0$ satysfying following condition
throughout pulsar's entire lifetime: 
\begin{equation}
E_0 = \min \{\zeta \cdot E_{\rm min}, \,E_{\rm W}, \,E_{\rm max}\}, 
 \label{e4}
\end{equation}
where $\zeta > 1$.

The best choice for the value of the parameter $\zeta$ was made a posteriori, to
reproduce the empirical trend of \Lg~ for the seven \cgro pulsars with similar accuracy
as equation (\ref{e2a}) does in Fig.2. We found that the range $2 \simless \zeta \simless 5$ fulfills
this requirement. All results presented below are for $\zeta = 2.5$. 
For $E_{\rm min}$ we preferred to take the numerically obtained values whenever they differred
from the analytical approximation 
(the analytical formulae for $E_{\rm W},\, E_{\rm max}$, and $E_{\rm min}$ are taken
from Rudak \& Ritter 1994). 
We found that the former are consistently smaller by a factor
of $\sim 1.5$ (in most cases) than the latter.

\begin{figure}
%\centering
%\vspace{14 cm}
\begin{center}
\leavevmode
\epsfxsize=8.4cm \epsfbox{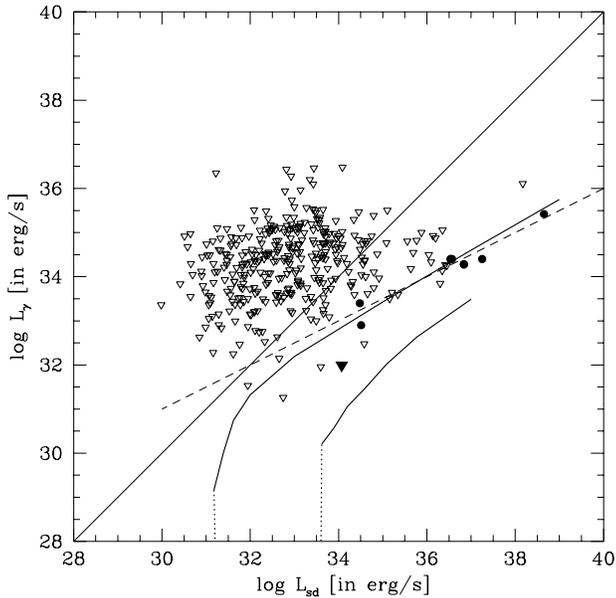}
\end{center}
\caption[pozycja spisu]{`Total' gamma-ray luminosity is plotted against 
spin-down luminosity \edot~for seven \cgro pulsars (filled dots).
Open triangles are the \egret upper limits as given by Nel et al.1996 for 350 objects, including
seven millisecond pulsars.
Filled triangle indicates position of J0437-4715 (after Fierro et al.1995).
The dashed line corresponds to 
$L_\gamma = 10^{16} \cdot L_{\rm sd}^{1/2}$ (see Section 2). The upper solid
curve shows the evolutionary track of a pulsar with $B_{\rm pc} = 10^{12}$G, calculated
according to eqs. (\ref{e3}) and (\ref{e4}). The lower solid curve shows the evolutionary track for
a millisecond pulsar with $B_{\rm pc} = 10^9$G.
The evolutionary tracks end up when the objects reach their `death points' in the \edot~space
(marked with dotted lines). 
}
\label{fig:fig3}
 \end{figure}

The developement of cascades was followed by means of numerical simulations described
in DH82.
Calculations of number of pairs $n_\pm$ were performed 
with numerical simulations after choosing $P$ and $B$, and setting the primary electron energy
$E_0$ according to equation (\ref{e4}). The normalization constant in our model of \Lg~
(equation \ref{e3}) was determined
by fitting numerical results to the seven detections (the last column of Table 1.). 
Then we calculated two evolutionary tracks in the \Lg~-- \edot~ space for representatives of
the classical pulsars (with typical magnetic field strength $B \sim 10^{12}$G), as well as
of the millisecond pulsars ($B \sim 10^9$G). Both tracks are shown in Fig.\ref{fig:fig3} as
solid curves. The upper curve ($B \sim 10^{12}$G) starts at $L_{\rm sd} \simeq 10^{39}\ergs$, nearby Crab,
and down to $L_{\rm sd} \sim 10^{34}\ergs$ ~it roughly follows the dashed line, which
depicts the relation $L_\gamma = 10^{16} \cdot L_{\rm sd}^{1/2}$ from Section 2. As \edot~ decreases, 
our exemplary classical pulsar enters a region where proximity to pulsar's death line
becomes important. The number of created pairs
$n_\pm$ declines constantly as the pulsar slows down, and it starts to decrease dramatically 
when $E_{\rm W}$ falls below $\zeta \, E_{\rm min}$, affecting thus $E_0$ in equation (\ref{e4}).
 
 At the point where
$E_{\rm W} = E_{\rm min}$, the creation of pairs ceases ($n_\pm = 0$) -- the pulsar reaches its death line.
The exemplary millisecond pulsar ($B_{\rm pc} = 10^9$G) follows the lower solid curve in Fig.\ref{fig:fig3}.
It starts at $L_{\rm sd} \simeq 10^{37}\ergs$, corresponding to initial period of one millisecond.
Unlike the former classical pulsar, it encounters different death line at $L_{\rm sd} \sim 10^{33.5}\ergs$,
due to an equality $E_{\rm max} = E_{\rm min}$. 
The efficiency of pair creation for the millisecond
pulsar throughout its lifetime, is significantly lower than for the classical pulsar. 
Low strength of magnetic field plays a decisive role here, and it cannot be compensated for by faster
rotation (and therefore -- by smaller curvature radii available).
The filled triangle in Fig.\ref{fig:fig3} indicates the upper limit for J0437-4715 
set from \egret observations (after Fierro et al.1995).  In addition, 350 \egret upper limits 
(Nel et al. 1996),
including seven limits for millisecond pulsars, are shown for comparison.
Apart from J0437-4715, four objects with \egret upper limits only are placed clearly below 
the upper evolutionary track. These are
(from right to left): B1046-58, B0656+14, B1929+10, and B0950+08. If the proposed model is correct, these objects
should be the best candidates for detection in gamma-rays (but not necessarily in the
\egret energy range).
B1951+32 and B1509-58, present in the data of Nel et al.~(1996), have been replaced with detections (filled dots)
by \egret and \osse, respectively.

%\newpage
 
\begin{figure}
%\centering
%\vspace{14 cm}
\begin{center}
\leavevmode
\epsfxsize=8.4cm \epsfbox{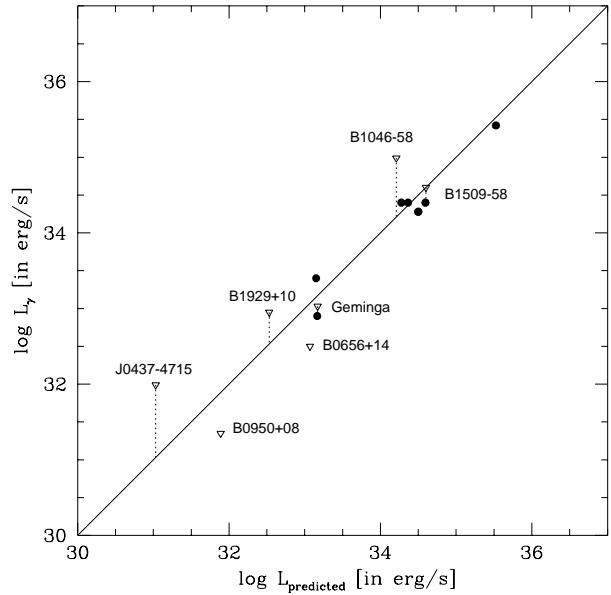}
\end{center}
\caption[pozycja spisu]{
`Total' gamma-ray luminosity is plotted against 
predicted luminosity, calculated
according to eqs. (\ref{e3}) and (\ref{e4}), for seven \cgro pulsars (filled dots).
Open triangles denote combined \osse, \comptel and \egret upper limits (wherever available).
The continuous diagonal line corresponds to a perfect agreement between
predictions and observations. 
} 
\label{fig:fig4}
 \end{figure}
 
The comparison of how our model reproduces \Lg~ for the seven \cgro pulsars, along with combined
upper limits from \egret, \comptel, and \osse 
(from Thompson et al.1994, Fierro et al.1995, Schroeder et al.1995, and Carrami{\~n}ana et al.1995),
wherever available, is shown in Fig.\ref{fig:fig4}.
In the case of Geminga and B1509-58, the model overestimates 
the observed \Lg~ by a factor of $\sim 2$. However, existing
upper limits from \comptel and \egret, respectively, improve the agreement. 
The upper limit for J0437-4715, based on \egret only, is one order of magnitude
above the predicted value of \Lg. For four other objects (B1046-58, B0656+14, B1929+10, and B0950+08)
the stringent upper limits
from \egret differ from the model predictions by no more than a factor of $\sim 3$.
Moreover,
upper limits from \comptel on B1046-58 and B1929+10 place them on a safe 
side of the diagonal line of prefect agreement in Fig.\ref{fig:fig4}. 
There is no information available about B0950+08 and B0656+14 from any \comptel observations.
 
\section{SUMMARY}

We have proposed a semi-phenomenological model of gamma-ray emission from pulsars, which is
based on polar cap activity triggered by
primary electrons. The energy of electrons is only a few times higher than the threshold energy
required to induce pair creation in the presence of a dipolar magnetic field, with other
restrictions applied when necessary.    
Electromagnetic cascades induced via curvature radiation
were treated in the same way as described by DH82.
The important ingredient of the model is the assumption that secondary particles, produced
in cascades due to one-photon absorption, contribute to overall gamma-ray emission
similarly as primary electrons. 

The model was confronted with the gamma-ray luminosities for seven pulsars inferred from
available data from \osse, \comptel, and \egret experiments. We find that the model is
consistent with the existing data. Moreover,  
the model does not lead
to any violence of energetics for pulsars with low spin-down luminosity \edot, -- the predicted 
gamma-ray luminosity \Lg~never reaches \edot. It avoids, therefore, the problem
of `an empirical gamma-ray death-line' as raised by Arons (1996). 
We also used the \egret archive of 350 upper limits 
along with \osse, and \comptel upper limits (published for 15, and 18 pulsars, respectively),
to find likely restrictions on the model. We have used some updates with respect
to the data used by Nel et al.(1996) in their analysis. 
The \egret upper limit for B1951+32 was replaced with 
its detections by \egret (Ramanamurthy et al. 1995), and \comptel (Kuiper et al.1996a).
Similarly, the \egret upper limit for B1509-58 was replaced with \osse detection (Schroeder et al.1995).
In the case of B1929+10 the model distance  of 170~pc
was replaced with 250~pc (see Yancopoulos, Hamilton \& Helfand~1994 for detailed arguments).

For a fixed value of
\edot, the predicted \Lg~depends rather weakly on magnetic field strength $B$ as long as
$10^{11}{\rm G} \simless B \simless 10^{13}{\rm G}$ (Dyks 1997), especially for pulsars
with $L_{\rm sd} \simgreat 10^{34}\ergs$. That is why all evolutionary tracks calculated for
high values of $B$  converge roughly
to the asymptotic relation
of equation (\ref{e2a}) for $L_{\rm sd} \simgreat 10^{34}\ergs$ (dashed line in Fig.3).
Only as $B$ enters the domain of millisecond pulsars
($\sim 10^8 - 10^9{\rm G}$), \Lg~ drops significantly. Therefore, all millisecond pulsars, including
J0437-4715, are expected to be very weak gamma-ray emitters regardless of their \edot, and
their \egret upper limits alone are still one order of magnitude
above our predictions for their gamma-ray luminosities. Qualitatively quite similar behaviour of millisecond
pulsars, though for different physical reasons,
results from the model of Dermer \& Sturner 1994,
which was  used explicitely in the context of millisecond pulsars
by Sturner \& Dermer 1994. For their luminosity $L_{SD94} = 1.1 \times 10^{10} B^{3/2} P^{-3}\ergs$ 
of gamma-rays beamed into a solid angle 
$\Omega_\gamma \approx 1.5 \times 10^{-3} P^{-1}$sr
(eqs.2 and 3 of Sturner \& Dermer 1994, respectively), the apparent gamma-ray luminosity 
for 1 sr can be expressed as
\begin{equation}
L_\gamma = \Omega_\gamma^{-1}L_{SD94} \approx 1.2 \times 10^9 \, B^{1/2} L_{\rm sd}^{1/2}\ergs, 
 \label{e7}
\end{equation}
and accordingly
for two pulsars with $B = 10^9{\rm G}$ and $10^{12}{\rm G}$ 
but identical spin-down luminosity 
\edot~ the former object will be placed below the latter one in a diagram like Fig.3.
The \egret upper limit for J0437-4715 ($\sim 10^{32}\ergs$) is well above  
$1.6 \times 10^{30}\ergs$ resulting from equation (\ref{e7}).

Out of~ several objects in the analysis of Nel et al.(1996), with uncomfortably low \egret
upper limits, and contradicting thus the models they discuss, only two are
left as a potential threat to our model: B0950+08, and B0656+14. 
The \egret limits for these two sources are 
too low to be accommodated by the model.
It is encouraging, however, that B0656+14 was reported as a possible
\egret source (Ramanamurthy et al. 1996).
Both pulsars were on the priority list of \comptel
but with low ranks, and no results are available for them so far. There are no \osse limits
available for B0656+14 either.
Definitely, 
B0656+14 deserves more attention as a promising target for gamma-ray experiments below
the energy range of \egret. Its parameters are very similar to those of Geminga and B1055-58.
Moreover, all three pulsars are strong X-ray emitters, and are thought to be the best candidates for
initial cooling ({\"O}gelman 1995, Becker \& Tr{\"u}mper 1997). On the other hand, 
the combined upper limits available for B1046-58,
and B1929+10 (\egret, \comptel, \osse in both cases) don't rule out our model.

The upper limits for energy fluxes adopted from Nel et al.(1996), and used also in this work 
require a word of comment. They were inferred from upper limits for photon fluxes 
under an assumption that
all photon spectra above 100~MeV have a spectral index $\gamma$, obeying a trend derived by
Thompson et al.(1994) from five \egret pulsars: 
\begin{equation}
\gamma = 0.33 \log \tau - 3.08, 
 \label{e5}
\end{equation}
where the characteristic
age of pulsars $\tau = P/2\dot P$ is expressed in years. The trend is based essentially on  
the Crab pulsar ($\gamma=2.16,\, \tau=1.3 \times 10^3$) on one side,
and on B1055-52 ($\gamma=1.18,\, \tau=5.3 \times 10^5$) on the other side. With the new determination
of the spectral slope for B1055-58, $\gamma=1.59$ (Fierro 1995), the prescription 
for $\gamma(\tau)$ looks questionable. As a consequence, upper limits for \Lg~ derived for old pulsars,
especially for millisecond pulsars, might be somewhat tighter. That would put models discussed by 
Nel et al.(1996) into even deeper trouble, whereas the model we propose still remains intact.

Pulsars from the database
of Taylor, Manchester \& Lyne (1993) extended by Taylor et al. (1995), plus Geminga,
arranged in a traditional ranking,
based just on spin-down fluxes $L_{\rm sd}/D^2$,
start with Crab and five other gamma-ray pulsars. But then, there is a wide gap
(of no gamma-ray detections) before the seventh
gamma-ray pulsar, B1055-52, emerges as No.33. 
The gap contains
several millisecond pulsars, with their flagship J0437-4715 taking very
high overall position -- No.7. 

Our ranking of top 30 candidates for gamma-ray emission,
arranged by a predicted flux resulting from equations (\ref{e3}) and (\ref{e4}),
\begin{equation}
f_\gamma = C \cdot {n_\pm \cdot E_0 \cdot L_{\rm sd}^{1/2}\over D^2},
 \label{e6}
\end{equation}
is presented in Table~2.
The pulsar database of Taylor et al.(1993, 1995) ordered by $f_\gamma$
starts now with Vela, then goes Crab and Geminga.
B1055-52 advances by 13 positions to No.20. The millisecond
pulsars (from the gap), including J0437-4715, disappear from the list of `Top 30'. 

\newpage

\begin{table}
\quad\quad\quad\label{t:par2}
\caption{Our ranking of top $30$ gamma-ray candidates.
The seven \cgro detections are marked with $\gamma$.} 
\vskip 1mm
\begin{tabular}{|r|c|c|c|c|c|}
\hline
 & PSR B & PSR J & & $D$ & $L_\gamma/D^2$ \\
 &     &       & & [kpc] & $[\ergscm2]$ \\
\hline
 1&0833$-$45   &0835$-$4510   &$\gamma$&0.50& 0.1302E-07  \\
 2&0531$+$21   &0534$+$2200   &$\gamma$&2.00& 0.8963E-08  \\
 3&          &0633$+$1746    &$\gamma$&0.15& 0.7308E-08  \\
 4&1706$-$44   &1709$-$4428   &$\gamma$&1.82& 0.6430E-09  \\
 5&1929$+$10   &1932$+$1059   &&0.25& 0.5678E-09  \\
 6&0950$+$08   &0953$+$0755   &&0.12& 0.5646E-09  \\
 7&1951$+$32   &1952$+$3252   &$\gamma$&2.50& 0.4108E-09  \\
 8&0656$+$14   &0659$+$1414   &&0.76& 0.2130E-09  \\
 9&1509$-$58   &1513$-$5908   &$\gamma$&4.40& 0.2080E-09  \\
10&1046$-$58   &1048$-$5832   &&2.98& 0.1904E-09  \\
11&          &2043$+$2740   &&1.13& 0.1653E-09  \\
12&1823$-$13   &1826$-$1334   &&4.12& 0.1239E-09  \\
13&1800$-$21   &1803$-$2137   &&3.94& 0.1193E-09  \\
14&0740$-$28   &0742$-$2822   &&1.89& 0.1143E-09  \\
15&0114$+$58   &0117$+$5914   &&2.14& 0.1136E-09  \\
16&1757$-$24   &1801$-$2451   &&4.61& 0.9188E-10  \\
17&          &1908$+$0734   &&0.58& 0.9129E-10  \\
18&1727$-$33   &1730$-$3350   &&4.24& 0.7281E-10  \\
19&          &0538$+$2817   &&1.77& 0.6567E-10  \\
20&1055$-$52   &1057$-$5226   &$\gamma$&1.53& 0.6346E-10  \\
21&0823$+$26   &0826$+$2637   &&0.38& 0.6209E-10  \\
22&          &1918$+$1541   &&0.68& 0.5512E-10  \\
23&0355$+$54   &0358$+$5413   &&2.07& 0.4791E-10  \\
24&1853$+$01   &1856$+$0113   &&3.30& 0.4525E-10  \\
25&0450$+$55   &0454$+$5543   &&0.79& 0.4274E-10  \\
26&1822$-$09   &1825$-$0935   &&1.01& 0.4079E-10  \\
27&1702$-$19   &1705$-$1906   &&1.18& 0.3806E-10  \\
28&1133$+$16   &1136$+$1551   &&0.27& 0.2123E-10  \\
29&0906$-$17   &0908$-$1739   &&0.63& 0.1433E-10  \\
30&1451$-$68   &1456$-$6843   &&0.45& 0.8666E-11  \\
\hline
\end{tabular}
%}
%}
%}
\end{table}

\section*{ACKNOWLEDGEMENTS}
Numerical code used in this work is based on a code originally developed and kindly provided by A.K.~Harding.
This research has been financed by the KBN grant 2P03D.009.11. We acknowledge useful remarks on the
typescript from T.~Bulik. We thank the anonymous referee for comments and useful suggestions.

\section*{ADDENDUM}
Already after submitting our paper for publication Lucien Kuiper pointed to us 
that the COMPTEL group has found indications for a signal from B0656+14 in the energy interval
of $10 - 30\,$MeV
(Hermsen, W., et al., 1997, Proceedings 2nd INTEGRAL Workshop, ESA SP-382, 287).

\label{lastpage}

\end{document}